\documentclass[conference]{IEEEtran}
\ifCLASSINFOpdf
  \usepackage[pdftex]{graphicx}
  \DeclareGraphicsExtensions{.pdf,.jpeg,.png}
\else
\fi
\usepackage{amsfonts}

\usepackage{hyperref}

\usepackage{siunitx}

\usepackage{booktabs}

\usepackage[%
]{glossaries}

\makeglossaries
\newacronym{roc}{ROC}{Receiver Operating Characteristic}
\newacronym{eer}{EER}{Equal Error Rate}
\newacronym{auc}{AUC}{Area Under Curve}
\newacronym{rff}{RFF}{Radio Frequency Fingerprinting}
\newacronym{em}{EM}{Electromagnetic}
\newacronym{sdr}{SDR}{Software Defined Radio}
\newacronym{rf}{RF}{Radio Frequency}
\newacronym{per}{PER}{Packet Error Rate}
\newacronym{fpr}{FPR}{False Positive Ratio}
\newacronym{tpr}{TPR}{True Positive Ratio}
\newacronym{fnr}{FNR}{False Negative Ratio}
\newacronym{snr}{SNR}{Signal to Noise Ratio}
\newacronym{sjr}{SJR}{Signal to Jamming Ratio}
\newacronym{gnss}{GNSS}{Global Navigation Satellite System}
\newacronym{uav}{UAV}{Uncrewed Aerial Vehicle}
\newacronym{leo}{LEO}{Low Earth Orbit}
\newacronym{sbc}{SBC}{Single Board Computer}
\newacronym{gpu}{GPU}{Graphical Processing Unit}
\newacronym{iq}{IQ}{In-phase Quadrature}
\newacronym{sdpsk}{SDPSK}{Symmetrical Differential Phase Shift Keying}

\begin{document}
\title{OrbID: Identifying Orbcomm Satellite RF Fingerprints}


% \author{Anonymous Authors}

\author{\IEEEauthorblockN{Cédric Solenthaler}
	\IEEEauthorblockA{ETH Zurich\\
		csolenthaler@ethz.ch}
	\and
	\IEEEauthorblockN{Joshua Smailes}
	\IEEEauthorblockA{University of Oxford\\
		joshua.smailes@cs.ox.ac.uk}
	\and
	\IEEEauthorblockN{Martin Strohmeier}
	\IEEEauthorblockA{armasuisse Science + Technology\\
		martin.strohmeier@ar.admin.ch}}
	

\IEEEoverridecommandlockouts
\makeatletter\def\@IEEEpubidpullup{6.5\baselineskip}\makeatother

\IEEEpubid{\parbox{\columnwidth}{
		{\fontsize{7.5}{7.5}\selectfont Workshop on Security of Space and Satellite Systems (SpaceSec) 2025 \\
		24 February 2025, San Diego, CA, USA\\
        ISBN 979-8-9919276-1-1 \\
		https://dx.doi.org/10.14722/spacesec.2025.23031\\
        www.ndss-symposium.org}
		%https://spacesec.info/
}
\hspace{\columnsep}\makebox[\columnwidth]{}}

\maketitle

% load generated evaluation results
\newcommand\name{Model_20241210_ORBCOMM_triplet}
\newcommand\modelsizetrainable{1091348}
\newcommand\modelsizetrainablesize{4.2 MB}
\newcommand\overallauc{0.53}
\newcommand\overalleer{0.48}
\newcommand\eeraccuracy{0.52}
\newcommand\multianchorauc{{0: 0.5914752865487408, 1: 0.6091678650242118, 2: 0.6189394767305535, 3: 0.6159994436520593, 4: 0.6117485255903142}}
\newcommand\multianchormaxauc{0.62}
\newcommand\multianchormineer{0.44}
\newcommand\multianchormineeraccuracy{0.54}
\newcommand\spoofingauc{0.98}

\newcommand\totalsize{14.4 GB}
\newcommand\datasettotalsamples{8992474}
\newcommand\datasettotalsampleshours{50.0}
\newcommand\datasetaverageSNR{8.1\,dB}

\newcommand\iridiumauc{0.576}
\newcommand\iridiumeer{0.448}

% Anonymize or deanonymize locations:
\newcommand\locSG{St. Gallen}
\newcommand\locZH{Zurich}
% \newcommand\locSG{\textit{Location 1}}
% \newcommand\locZH{\textit{Location 2}}


\begin{abstract} 
An increase in availability of Software Defined Radios (SDRs) has caused a dramatic shift in the threat landscape of legacy satellite systems, opening them up to easy spoofing attacks by low-budget adversaries.
Physical-layer authentication methods can help improve the security of these systems by providing additional validation without modifying the space segment.
This paper extends previous research on Radio Frequency Fingerprinting (RFF) of satellite communication to the Orbcomm satellite formation.
The GPS and Iridium constellations are already well covered in prior research, but the feasibility of transferring techniques to other formations has not yet been examined, and raises previously undiscussed challenges.

In this paper, we collect a novel dataset containing \datasettotalsamples{} packets from the Orbcom satellite constellation using different SDRs and locations.
We use this dataset to train RFF systems based on convolutional neural networks.
We achieve an ROC AUC score of \overallauc{} when distinguishing different satellites within the constellation, and \spoofingauc{} when distinguishing legitimate satellites from SDRs in a spoofing scenario.
We also demonstrate the possibility of mixing datasets using different SDRs in different physical locations.

\end{abstract}


% \fi
\IEEEpeerreviewmaketitle

\section{Introduction}
\subsection{Motivation}

Over the past decades \glspl*{sdr} have proliferated and in the process have invalidated previously sensible wireless threat model assumptions.
Nowadays, commonly available and cheap hardware enables attacks that were, just a few decades prior, firmly in the domain of nation state actors.
In fact, many such wireless attacks are now occurring in the wild, such as GPS spoofing~\cite{LiveGPSSpoofingTrackerMap},
and recent research even demonstrated arbitrary code execution through wireless signal injection~\cite{VSATInjection}.
While these developments are likely to precipitate improved wireless security in the future,  some systems, such as satellite constellations,
cannot simply be replaced due to accessibility, legacy, or cost considerations.

In the domain of satellite communication security, security measures that rely on the physical layer and are deployable purely in the ground segment provide valuable tools to improve the security of legacy systems.
Physical-layer identification techniques have shown promising results for satellite identification~\cite{WatchThisSpace, 3DConvRFFIridium, PastAI}. These works tend to focus on the GPS and Iridium constellations due to the ease and ubiquity of receiving signals from these satellites using off-the-shelf hardware and open-source software.

In this paper, we extend existing satellite fingerprinting techniques to work with the Orbcomm satellite constellation. These satellites present a typical example of the legacy systems that RFF is suited to securing.
The Orbcomm constellation brings a number of unique challenges not present in the Iridium system:
\begin{itemize}
    \item Communication channels are low-bandwidth and differ between satellites;
    \item Small frequency spacing between channels prevents high levels of oversampling;
    \item Messages contain identifying information which, if improperly masked, could allow the model to cut corners.
\end{itemize}

\subsection{Contributions}

In this paper we provide the first application of RF fingerprinting to the Orbcomm constellation.
In order to achieve this, we collected a dataset of \datasettotalsamples{} samples that
were collected from multiple locations and \glspl*{sdr}.
Using this dataset we demonstrate the effectiveness of RFF to authenticate communication, and separate legitimate communication from spoofing via attacker-controlled SDRs.
We also introduce a novel preprocessing technique to anonymize messages without reducing the number of samples available to the fingerprinting model.
Finally, we show the stability of the fingerprints over time, and evaluate transferability between SDRs.

% \section{Content}
% Background

\section{Background}

\begin{figure*}
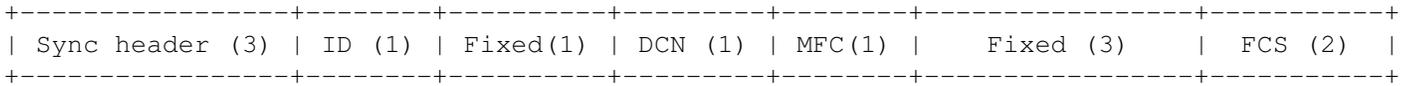

    \centering
\begin{verbatim}
+-----------------+--------+----------+---------+--------+-----------------+-----------+
| Sync header (3) | ID (1) | Fixed(1) | DCN (1) | MFC(1) |    Fixed (3)    |  FCS (2)  |
+-----------------+--------+----------+---------+--------+-----------------+-----------+
\end{verbatim}
    \caption{Structure of an Orbcomm synchronization packet.}
    \label{fig:sync-packet}
\end{figure*}

%This section presents the relevant background as well as prior work in the relevant area.

\subsection{Software Defined Radios}

\Glspl*{sdr} are radio devices that  digitize the received signal, offloading the bulk of (traditionally analog) signal processing to a computer.
\Glspl*{sdr} utilize IQ demodulation in order to downconvert the signal and split it into I (In-phase) and Q (Quadrature) components.
The I and Q signals, together with the original carrier frequency, can be used to fully reconstruct the received signal.
After the \acrshort*{iq} signals are sampled and digitized, they are commonly represented as a complex number where the real part corresponds to the in-phase component.

We focus on two commonly used off-the-shelf \Glspl*{sdr}.

\subsubsection{HackRF One}

The HackRF One is a Semi-Duplex \gls*{sdr} that covers a frequency range from \qtyrange{1}{6000}{\mega\hertz} with a maximal bandwidth of \qty{20}{\mega\hertz} and a resolution of 8 bits~\cite{greatscottgadgetsHackRFGreat}.

\subsubsection{RTL-SDR}
The RTL-SDR is a receive-only family of \gls*{sdr} devices that covers a frequency range from \qty{500}{\kilo\hertz} to \qty{1.75}{\giga\hertz} with a maximal bandwidth of around \qty{3}{\mega\hertz} and an effective resolution of 7 bits~\cite{rtlsdrAboutRTLSDR}.
The RTL-SDR is notable due to its very low price point, in the range of \num{30}~USD.

\subsection{Radio Frequency Fingerprinting}

\Gls*{rff} aims to identify an emitter based on imperfections in the \gls*{em} signal~\cite{ReviewRFF}. These imperfections stem from variations in the fabrication of the RF chain and are thus hard to avoid.
These variations can be used to discriminate signals from transmitters of the same model, and in particular can distinguish attacker-controlled transmitters from legitimate ones. This may be used to provide an extra security layer~\cite{ReviewRFF} against spoofing, which may be implemented independently of existing infrastructure.

There are two main approaches to \gls*{rff}, which focus on either the transient of the signal (the beginning of the signal as the transmitter powers up) or the steady state during normal transmission~\cite{ReviewRFF}.
Within each of these, some approaches use a classifier (in which all transmitters are known and messages are labeled) and some use similarity-based techniques (in which fingerprints are compared to a database)~\cite{DeepComplexNetworksForRFF}.

\Gls*{rff} is used in the literature to identify transmitters operating various protocols, including WiFi~\cite{DeepComplexNetworksForRFF,WirelessDeviceIdentificationwithRadiometricSignatures,DeepLearningCNNRFFIWLAN,MassiveExperimentalRFFStudy},
or aviation communication~\cite{DeepComplexNetworksForRFF, MassiveExperimentalRFFStudy}, as well as various satellite systems.
Among these the Iridium constellation~\cite{WatchThisSpace, PastAI, SatPrint} and \glspl*{gnss} ~\cite{GPSFingerprinting} stand out, in addition to some study of simulated CubeSats~\cite{RobustSatAntFingerprintingRNN}.

\subsection{Triplet Loss}
The triplet loss function~\cite{TripletLossOASIS} is a measure of similarity between sparse representations
%It is used for supervised learning, i.e. each example has a known label.
and is calculated from a triplet of embeddings.
It compares the distances from an anchor to a positive sample (same label as the anchor) and to a negative sample (different label).
The triplet loss characterizes how much closer the positive example is to the anchor than the negative example. Triplet loss encourages a model to map semantically close inputs to metrically close embeddings~\cite{1703.07737}.

Mathematically, the triplet loss $L$ is calculated as described in \autoref{eq:tripletloss}~\cite{TripletLossOASIS} where $p$ is the embedding of the anchor, $p_+$ is the embedding of the positive example, and $p_-$ is the embedding of the negative sample. The function $d(x, y)$ describes a distance between two embeddings $x$ and $y$.
The margin $m$ is a hyperparameter that describes the minimal desired difference between embeddings of different labels.
Due to the $\max(\cdot)$ term,
differences beyond the margin do not contribute to the loss anymore.

\begin{equation}
    L = \max (m +d(p, p_+) - d(p, p_-), 0) 
\label{eq:tripletloss}
\end{equation}

\subsection{Orbcomm}

The Orbcomm satellite constellation was established from 1995 to 2000~\cite{OrbcommExperience} and provides communication services. The space segment consists of 30 satellites that communicate in the \qtyrange{137}{150}{\mega\hertz}  (VHF) band~\cite{OrbcommExperience} with a right-hand circular polarization. Not all of these satellites appear to be functional~\cite{MultiConstellationSDRPositioning}. In practice, only 13 channels between  \qtyrange{137}{138}{\mega\hertz} are used~\cite{CarrierPhaseTrackingPositioningOrbcomm}, one of which is not a user channel but a gateway channel~\cite{OppertunisticNavigationOrbcommIridium}.

The signal is encoded with \gls*{sdpsk}, with 8~bits to a word and 600 words in a minor frame~\cite{OrbcommExperience}. Minor frames consist of 12 or 24 word-long packets. Each packet contains a two word Fletcher checksum~\cite{OrbcommProtocol}.
Orbcomm satellites transmit continuously~\cite{MultiConstellationSDRPositioning}
and with a transmission rate of \qty{4800}{bps}.

There are several different types of used packets. 
\autoref{fig:sync-packet} shows the structure of a synchronization packet that contains a 
synchronization header (0x65A8F9), the satellite ID, a Downlink Channel Number (DCN), a Minor Frame Counter (MFC), as well as a fletcher checksum (FCS) and four fixed words~\cite{OrbcommProtocol}.
Both the ID and the DCN field (and the checksum too since only ID and DCN change between different synchronization packets) can be used to infer the identity of the satellite.

\subsection{Related Work}

While there is an abundance of prior work~\cite{WatchThisSpace,3DConvRFFIridium,PastAI,DeepComplexNetworksForRFF,SatPrint,GPSFingerprinting,RobustSatAntFingerprintingRNN,RFFSpikingNN} using \gls*{rff} on satellite communication (real or simulated), to the best of our knowledge there is no prior work evaluating the Orbcomm constellation.
However, several papers do use Orbcomm to provide opportunistic positioning~\cite{MultiConstellationSDRPositioning,CarrierPhaseTrackingPositioningOrbcomm,OppertunisticNavigationOrbcommIridium}.

Some previous \gls*{rff} works pre-process \acrshort*{iq} samples into an image (or multiple images) to produce heatmaps, which are used in classifiers pre-trained on image recognition models~\cite{3DConvRFFIridium,PastAI}.
Other works such as SatIQ~\cite{WatchThisSpace} perform 1D convolutions on the raw signal, similar to conventional signal processing. We use this approach in this paper.

Another important difference is into what space the model projects the input.
Some previous works use a classification approach~\cite{3DConvRFFIridium,PastAI,DeepLearningCNNRFFIWLAN}, whereas other works~\cite{WatchThisSpace} opt for a more extendable similarity-based approach, enabling new transmitters to be added without retraining.
The methodological difference to SatIQ~\cite{WatchThisSpace} lies in not using a synchronized signal as input for the proposed model and making more use of shared weights in processing.

% Threat Model

\section{Threat Model}

Satellite systems (especially GPS) are commonly spoofed in the wild~\cite{LiveGPSSpoofingTrackerMap}.
Cheap, transmit-capable \glspl*{sdr} are readily available (e.g. HackRF for around 300\,USD~\cite{greatscottgadgetsHackRFGreat}) and are more than sufficient to perform many attacks.
An attacker with reasonably low budget (2000\,USD) is able to overshadow satellite systems~\cite{SatelliteSpoofing} or even take over user ground stations~\cite{VSATInjection}.

% \subsection{The Adversary}

Thus, we can assume the adversary has the capability to overshadow the legitimate signal or replay signals from the transmitter. Due to the low power of legitimate satellite signals on the ground, the attacker is assumed to be able to overshadow a satellite despite the user potentially utilizing a highly directional antenna.
Alternatively the attacker might target a sidelobe, which has been shown to be practical for small aperture (user segment) satellite dishes~\cite{VSATInjection}.

Compared to prior work~\cite{PastAI}, we also consider non-terrestrial adversaries, because
even a low-budget attacker can use an \gls*{uav} to carry an \gls*{sdr} and a \gls*{sbc} above a receiver and thus fool systems based on detecting terrestrial links (e.g. via fading process~\cite{PastAI}).

The user is unable to distinguish attacker messages from legitimate ones on the logical layer (e.g. due to absent/broken cryptography or leaked key material).
Thus the user must attempt to authenticate the signal on the physical layer.

A sufficient test of this capability consists of distinguishing different transmitters of the same constellation. Previous work~\cite{WatchThisSpace} indicates that the performance against a replay attacker (over a wired link) is better than the performance in distinguishing legitimate transmitters.

% \subsection{The User}

Out of scope for the purpose of this work is an adversary capable of hijacking the actual satellite hardware, thus taking over the transmitter.
Neither is the adversary able to inject malicious samples into our training data.
Such a capability would enable the adversary to create a backdoor in the machine learning model~\cite{SatBackdooring}.

% Data Collection
\section{Data Collection}

\begin{figure}
    \centering
    \includegraphics[width=0.7\linewidth]{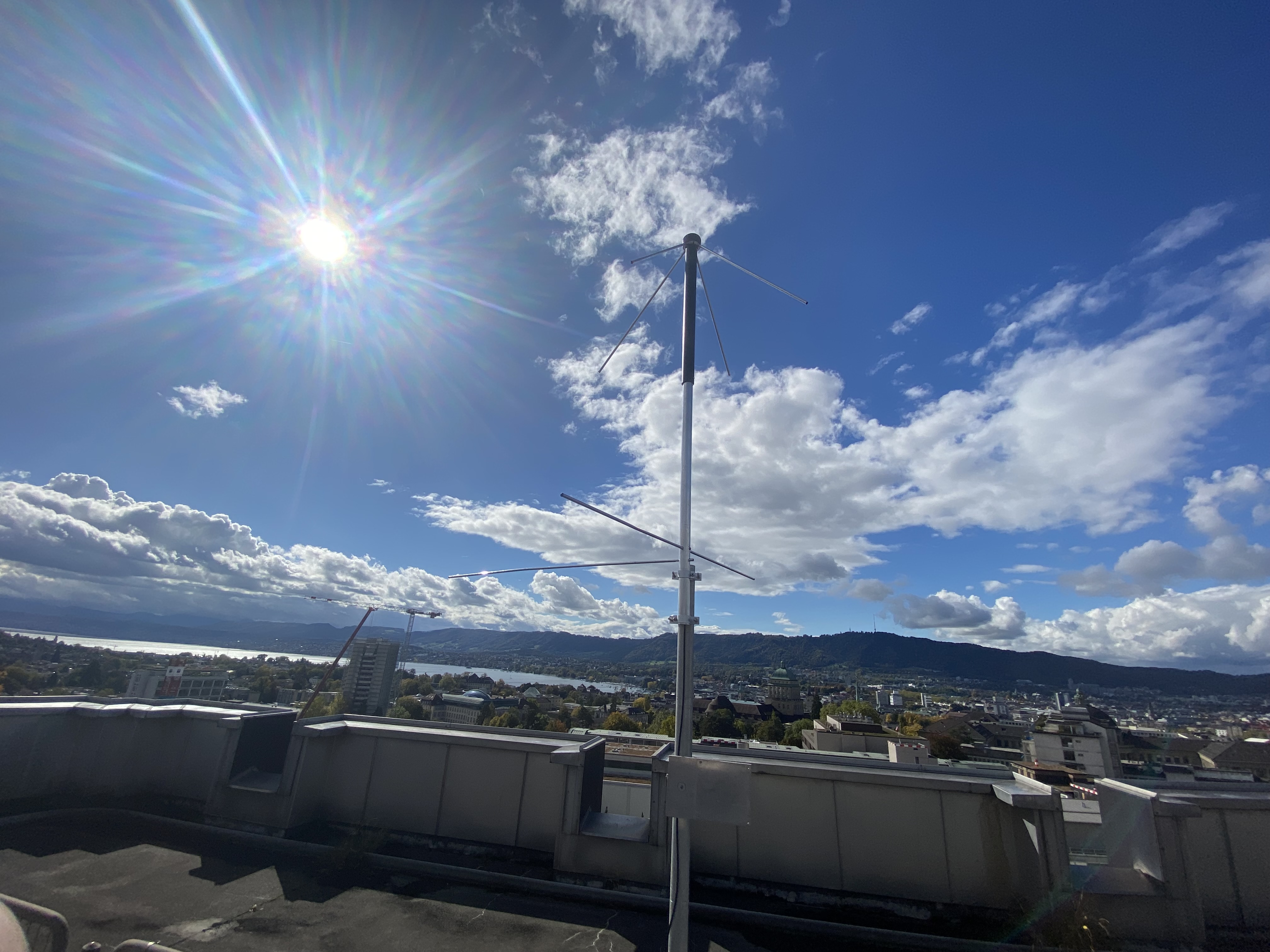}
    \caption{Turnstile antenna for the data acquisition system deployed on a rooftop in \locZH{}.}
    \label{fig:ZH-ant}
\end{figure}

Data was collected from two physical locations.
\autoref{fig:ZH-ant} shows a turnstile antenna designed for frequencies from \qtyrange{135}{152}{\mega\hertz}
and right-hand circular polarization~\cite{TA-1Kreuzdipol} deployed on a rooftop in \locZH{}.
It is connected via a coaxial cable to an RTL-SDR and a computer inside a weatherproof closet. Another turnstile antenna was deployed in \locSG{} and connected to a HackRF One. Preliminary experiments were also carried out with other antenna types (Dipole, V-Dipole, QHF) at \locSG{}. However, the turnstile antenna was found to deliver the best results.

% Software side
The computer runs a custom python script that acquires the samples.
The script uses the Soapy abstraction library to ingest the IQ samples. Much of the decoding logic is based on an open-source project~\cite{orbcomm-receiver}
that implements the decoding. It leverages ephemeris data for the satellite detection and prediction of the used channels, as well as the Doppler offset. If the receiver is able to synchronize to the incoming bitstream (i.e., estimate where word and packet boundaries are), all fletcher checksums are checked.
This is used to calculate the \gls*{per}.
Packets are only accepted into the dataset if the receiver validates its checksum.

\begin{figure}
    \centering
    \includegraphics[width=\linewidth]{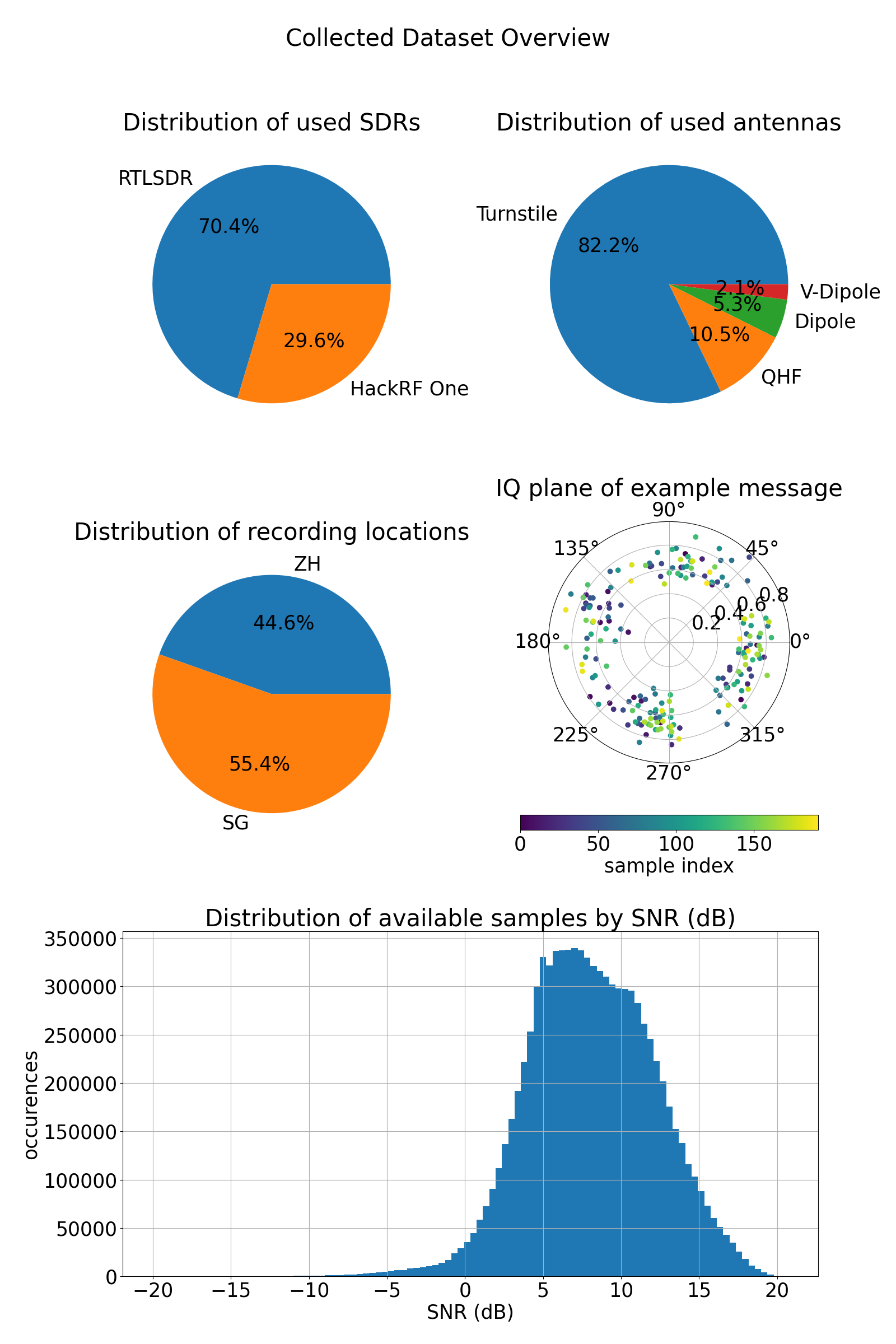}
    \caption{Breakdown of collected dataset}
    \label{fig:dataset-stats}
\end{figure}
The possible oversampling is limited by the separation of the channels to a value such that no neighboring channel is included.
That would be undesirable, because the presence or absence of neighboring bands could help an ML model to identify the transmitter, taking advantage of the variable spacing between downlink channels among the Orbcomm satellite fleet.
While the satellites could be programmed to transmit in steps of \qty{2.5}{kHz}~\cite{OrbcommProtocol}, only a few are actually used. Various online forums report frequencies in use by Orbcomm and the smallest channel separation found was \qty{20}{kHz}.
With a bitrate of \qty{4800}{bps} and a root raise cosine filter with $\alpha=0.4$~\cite{OrbcommProtocol}, the captured bandwidth should not be more than $20\text{kHz} - (2.4\text{kHz} * (1+\alpha)) = 16640 \text{Hz}$. To account for non-ideal filers, the signal is oversampled by a conservative factor of two.

The stored signal undergoes coarse frequency correction (Doppler shift prediction with ephemeris), fine frequency correction, but not timing nor phase recovery. The \acrshort*{iq} samples and associated metadata (\gls*{snr}, RX location, timestamp, used antenna, used \gls*{sdr}) are stored in a tfrecord file.
% Before training, the dataset was reshuffeled and the files merged into fewer, larger tfrecord files.
This data and the associated data collection code will be made available on publication.

% \subsection{Dataset}

Overall the data collection ran intermittently for 3 months (1.5 months thereof from two locations) and collected a total of \datasettotalsamples{} packets with \num{192} \acrshort*{iq} samples each.
This represents
% NOTE: This would cause compilation error: \qty{\datasettotalsampleshours{}}{h}
\qty{\datasettotalsampleshours}{h}
of continuous \acrshort*{iq} recordings.
The average \gls*{snr} is 
\datasetaverageSNR{}.
\autoref{fig:dataset-stats} summarizes the dataset and the distribution of the metadata of the collected samples.

% System Design
\section{System Design}

\subsection{Data Augmentation}

Because of concerns raised in previous research regarding the transferability of fingerprinting models across different \gls*{sdr} reboots~\cite{irfan2024reliabilityradiofrequencyfingerprintingPrePrint},
the data collected for this dataset stems from two different locations with two different physical \glspl*{sdr}.

\begin{figure}
    \centering
    \includegraphics[width=0.95\linewidth]{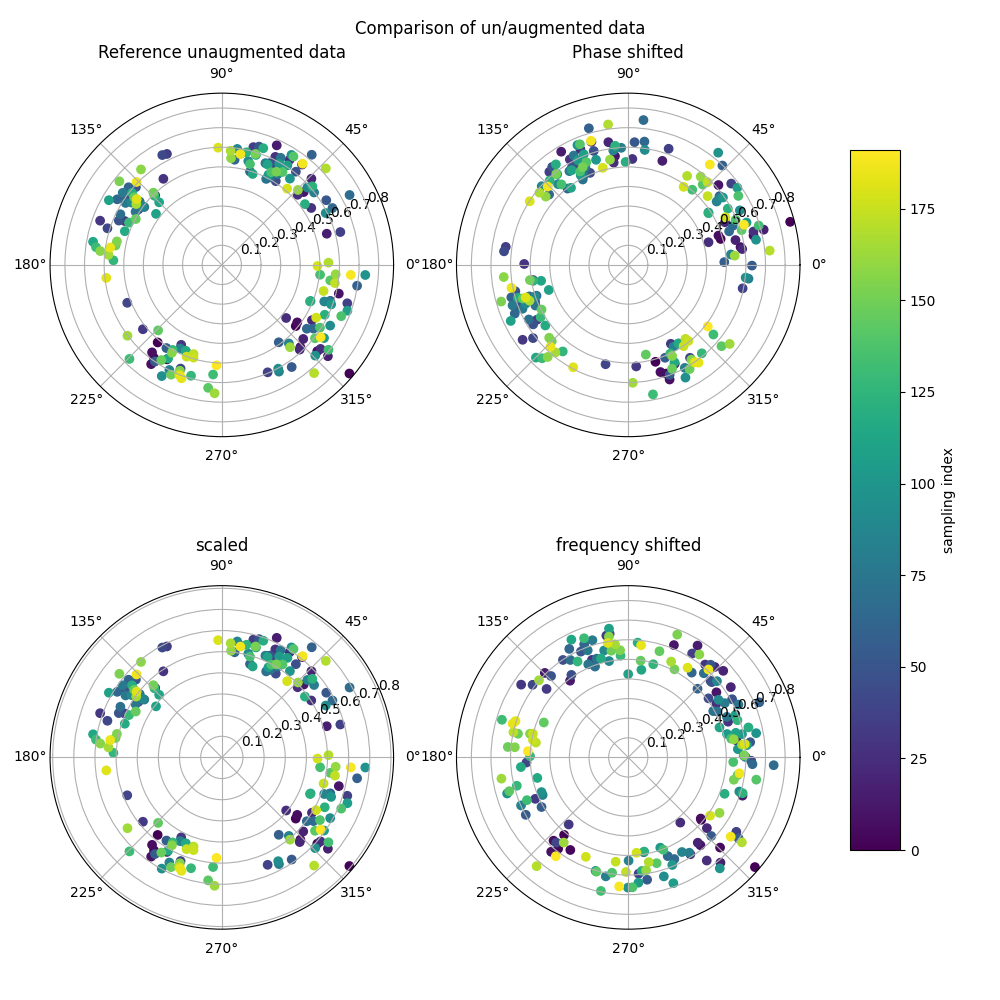}
    \caption{Illustration of data augmentation}
    \label{fig:data-augmentation}
\end{figure}

The data is then augmented in the dataset pipeline.
We apply the following operations: random phase shift ($\Delta\phi \in [0, 2\pi)$), random scaling ($\alpha \in [0.9, 1]$), and a random frequency shift ($\Delta f / f_{sampling} \in [-0.01, 0.01]$).
These transformations are illustrated in \autoref{fig:data-augmentation}.
The intent is to improve training and prevent overfitting.

\subsection{Machine Learning Model}

To provide an extendable system that can handle new categories without requiring extensive retraining, an encoder architecture is chosen that maps the received waveform (\acrshort*{iq} samples) into an embedding space.

\begin{figure*}[h]
    \centering
    \includegraphics[width=\textwidth]{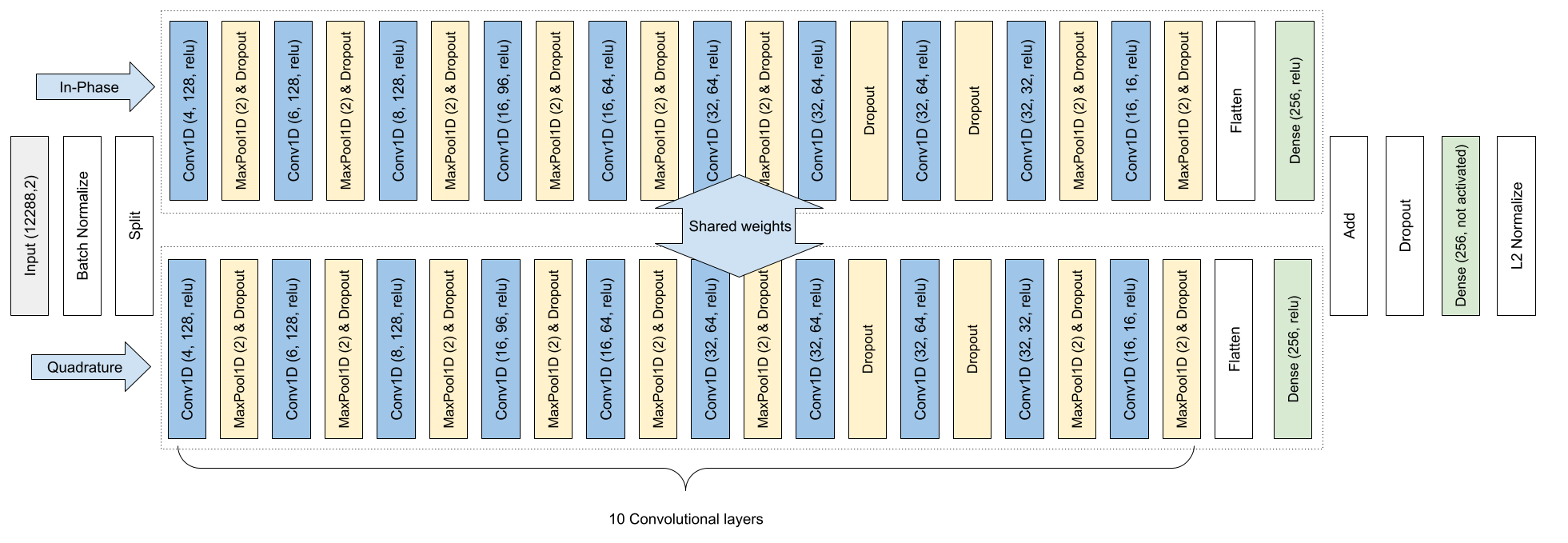}
    \caption{Proposed Model Architecture}
    \label{fig:model-arch}
\end{figure*}

\subsubsection{Loss Function}

The triplet loss with a semi-hard online batching strategy is used. Semi-hard batching considers all positive samples in a batch but only the hardest negative example and was found to provide more stable training~\cite{Schroff_2015}. The chosen distance metric is the L2 metric with a margin of 0.7, because this performed best in preliminary experiments.

\subsubsection{Data Preprocessing}

To improve model performance and increase the model size, the signal is interpolated during preprocessing.
This increases the input size of the model and enables bigger Conv + Maxpool
layers with more parameters. The following list
shows the preprocessing layers (and output tensor shapes) that are applied before the encoder.

\begin{enumerate}
    \item Input (Batch size, 192, 2)
    \item Interpolation (x64) to (Batch size, 12288, 2)
    \item Random Bit Transition Shuffling (Batch size, 12288, 2)
\end{enumerate}

Many Orbcomm packets contain fields with information that identifies the specific satellite (ID field or ephemeris data) or narrow down the possible satellites (e.g. information about used transmit frequencies)~\cite{OrbcommProtocol}, further preprocessing is needed in order to prevent  the model from learning to decode the Orbcomm signal.
The data portion of the signal is easily spoofed by the attacker, so we must ensure the model cannot extract any identifying information from it.

To achieve this, the signal is split into different bit transitions and shuffled randomly. Afterwards the different bit transitions are stitched back together.
This happens as a preprocessing layer inside the Keras model and is configured to only happen during training.
This step decreases the quality of the signal since it introduces discontinuities but is vital to prevent the model from learning logical-layer features that do not offer any protection against replay or spoofing attacks.

\subsubsection{Embedding Network}

For the embedding network, a convolutional neural network is chosen. It consists of two separate branches of Conv1D, MaxPool1D and Dropout layers that apply to the I and Q part of the signal, respectively. Using weight sharing on the convolution kernels, the model size is reduced. This weight sharing is motivated by the fact that the received \acrshort*{iq} signal can have any phase shift due to the wireless channel, and thus the \acrshort*{iq} parts can be swapped (and one of them multiplied by $-1$) if the phase shift is $\pi/2$.

After 10 convolutional layers, the processed I and Q signals pass through a dense layer that transform them to the desired embedding dimension. The final embedding dimension is 256. The signals from the I and Q branch are added and passed through a final dense layer. The full model architecture (excluding preprocessing layers) is shown in \autoref{fig:model-arch}.
The final model contains \modelsizetrainable{} trainable parameters (\modelsizetrainablesize{}).

\subsubsection{Training}

The model is implemented in Tensorflow with the help of Keras for many of the standard layers.
Non-standard layers include the triplet loss function (where code from the now-deprecated tensorflow-addons project is used) as well as the preprocessing layers (both the interpolation and the bit transition shuffling layers are custom).

Of the available samples, 27\% were used for testing and 9\% for validation. The plots in Section \ref{sec:evaluation} were generated with validation data if not otherwise mentioned.

Training was performed on an Nvidia A100 \gls*{gpu} with the adam optimizer~\cite{kingma2017adammethodstochasticoptimization}. A total of 25
epochs were trained over a time of 50 hours. The used batch size was 
450.
During training, every epoch that achieves a new best \gls*{auc} metric is saved, such that early stopping can be simulated. Actual early stopping was unhelpful as the \gls*{auc} metric did not evolve monotonically.

% Evaluation
\section{Evaluation}
\label{sec:evaluation}

\subsection{Accuracy}
\begin{figure}
    \centering
    \includegraphics[width=0.87\linewidth]{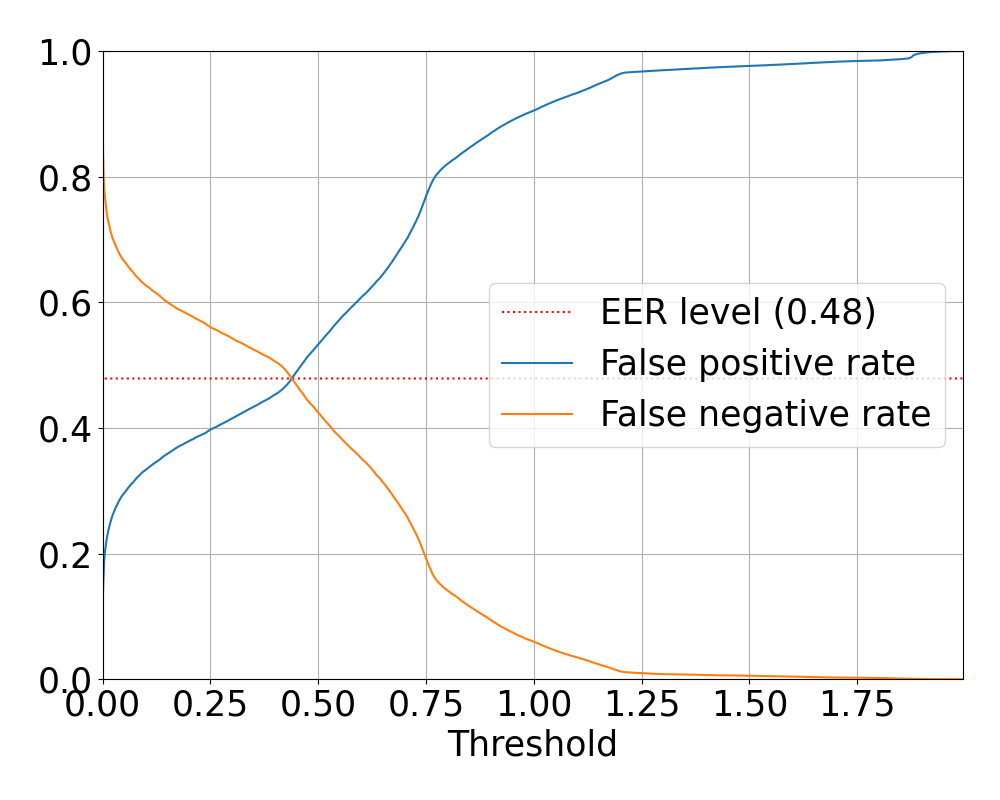}
    \caption{\gls*{eer} of model}
    \label{fig:roc-eer}
\end{figure}

We provide initial results of our approach. The \gls*{roc} characteristics are calculated by taking the pairwise distance metric between a batch of $N=1024$ validation examples and then varying the detection threshold. This method yields $(N^2 - N)/2$ binary comparisons per batch.
The threshold decides if the pairwise distance is classified as belonging to the same transmitter or not. The threshold is evaluated over the entire validation dataset. The overall \gls*{roc} \gls*{auc} is \overallauc{} with an \gls*{eer} of \overalleer{}. \autoref{fig:roc-eer} shows the \gls*{eer} of the \gls*{roc} curve.

\begin{figure}
    \centering
    \includegraphics[width=0.87\linewidth]{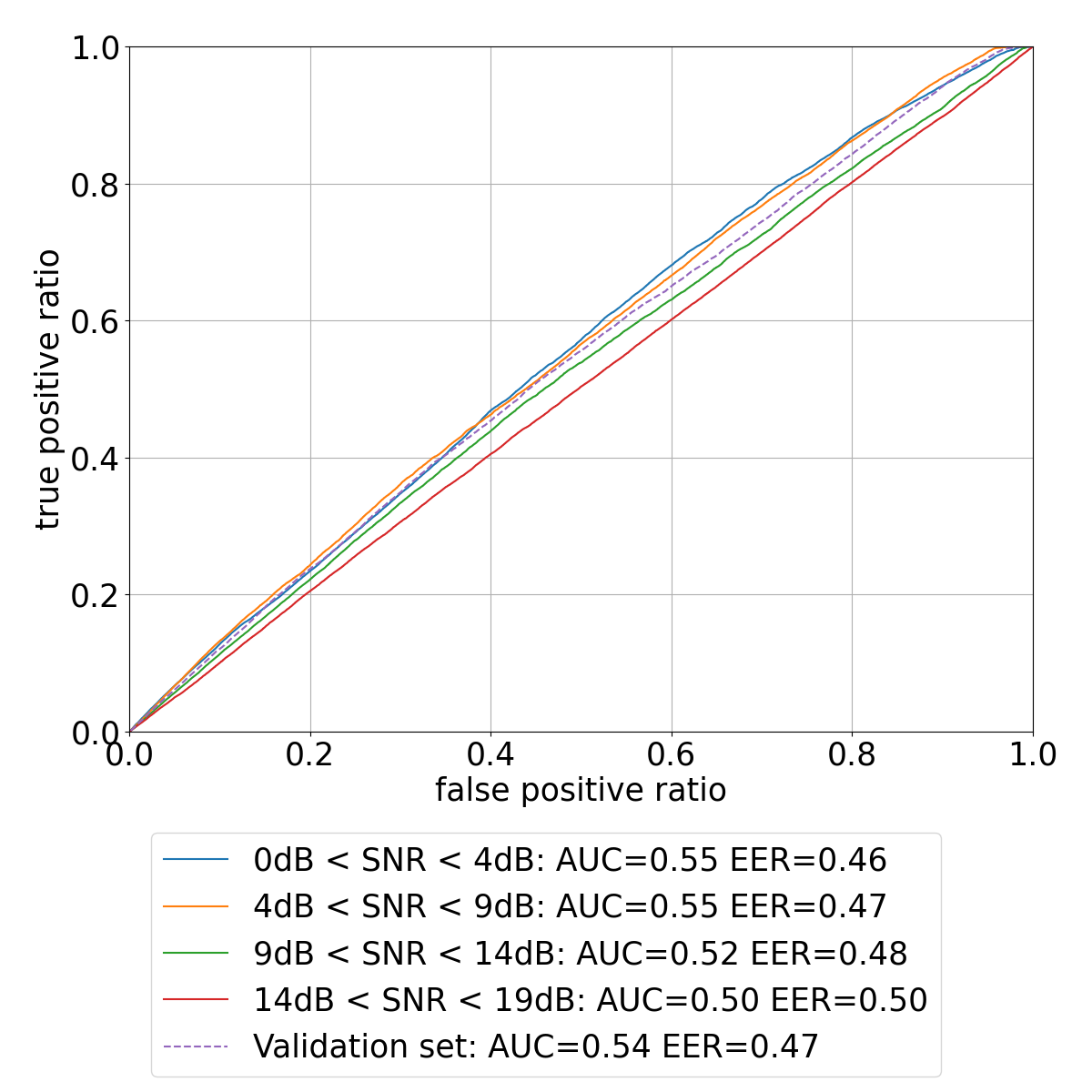}
    \caption{\gls*{roc} \gls*{auc} for different minimal \gls*{snr}}
    \label{fig:rocauc-snr}
\end{figure}

\autoref{fig:rocauc-snr} shows that the classification performance improves for higher \gls*{snr} until it decreases again. We postulate that our \gls*{snr} measurement is affected by interference which creates false entries with high \gls*{snr}.

\begin{figure}
    \centering
    \includegraphics[width=0.85\linewidth]{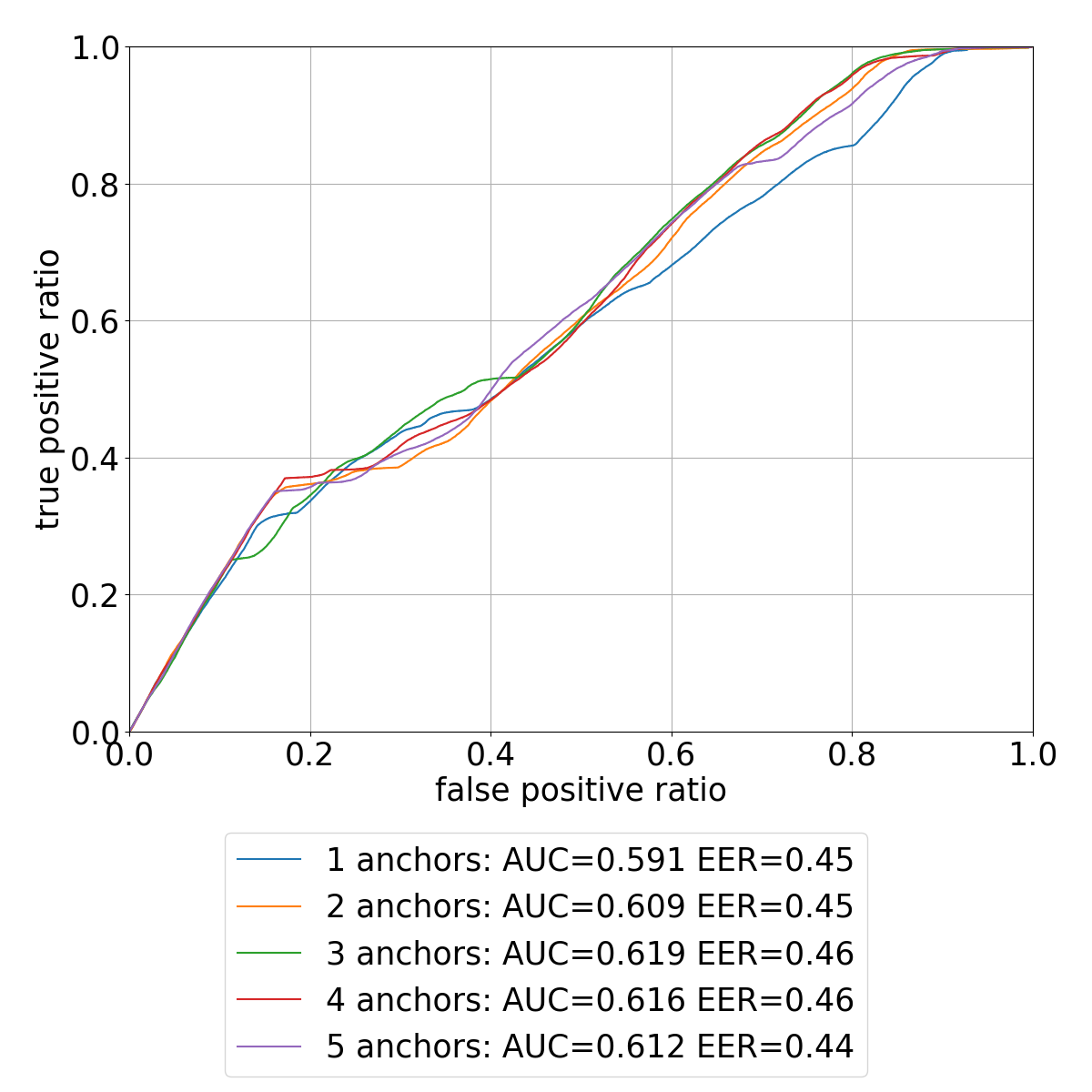}
    \caption{Anchored ROC}
    \label{fig:anchored-rocauc}
\end{figure}

To improve the performance of the classifier, a signal can not only be compared to a single anchor (a ground truth embedding from a known satellite) but to several anchors instead.
The anchors are randomly sampled and the mean distance between the tested embedding and the anchors. \autoref{fig:anchored-rocauc} shows performance for different amounts of anchors. This scenario is different from the baseline performance, in that the model is used to decide if a message belongs to the same transmitter as the anchor or not. In the previous baseline scenario, a sampled subset of all pairwise distances is used to decide whether two signals belong to the same satellite. 

The achievable baseline accuracy (at \gls*{eer}) is $1 - \text{EER} = \eeraccuracy{}$. When using three %several
anchors the \gls*{auc} is improved to
\multianchormaxauc{} and an accuracy of \multianchormineeraccuracy{}.

\subsection{Security}

\begin{figure}
    \centering
    \includegraphics[width=0.5\linewidth]{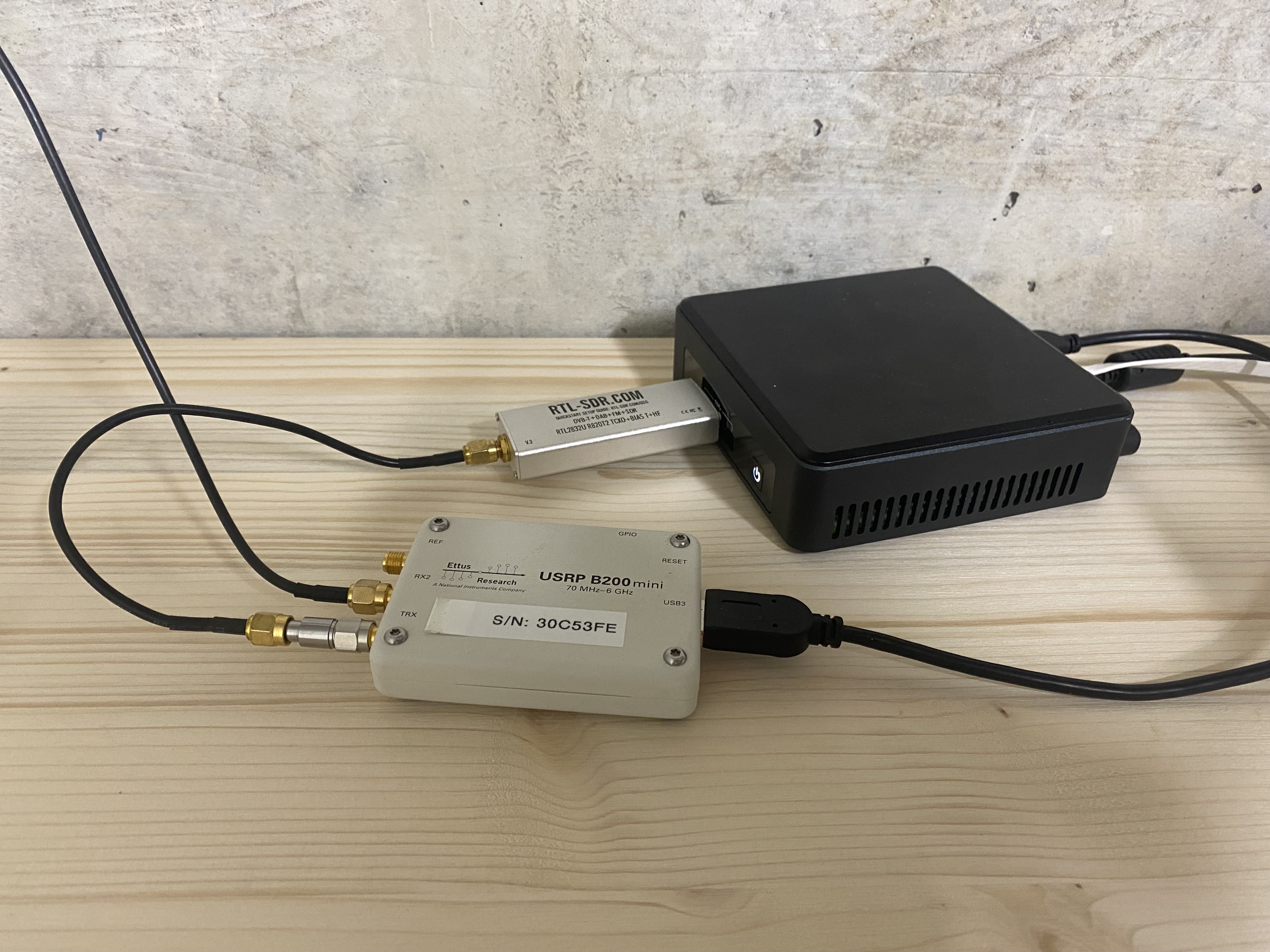}
    \caption{Signal Spoofing Setup}
    \label{fig:spoofing-setup}
\end{figure}

To assess the security of the proposed scheme, we collect a dataset of spoofed messages. We use a full-duplex USRP B200mini \gls*{sdr} to capture signals with a center frequency of \qty{136.9}{\mega\hertz} and a sampling rate of \qty{3}{Msps}. We use the same parameters to emit it on the transmit port, which is connected via a \qty{10}{\decibel} attenuator to the data collection system. \autoref{fig:spoofing-setup} shows the spoofing setup connected to the collection system.

\begin{figure}
    \centering
    \includegraphics[width=0.85\linewidth]{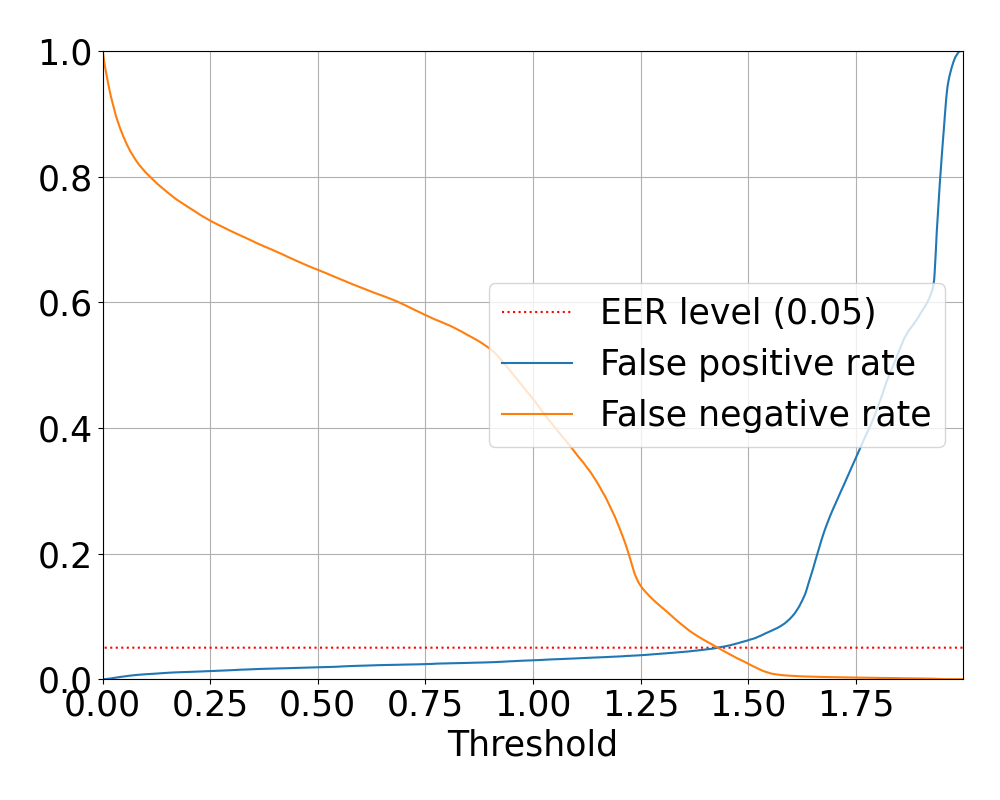}
    \caption{EER plot of model used to discriminate real and spoofed signals}
    \label{fig:spoofing-eer}
\end{figure}

For the analysis, 10000 real recordings and 10000 spoofed recordings are sampled from the datasets and transformed with the model into the embedding space.
For every satellite with entries in both sets, the distance between all spoofed and real embeddings is calculated, as well as all pairwise distances between embeddings of real samples. The \gls*{roc} for the binary discriminator (decision between spoofed and not spoofed) is \spoofingauc{}. \autoref{fig:spoofing-eer} shows the false positive / negative rates depending on the used threshold and the \gls*{eer} level.

\subsection{Deployability}

This subsection addresses some considerations that affect the deployability of the proposed system. The accuracies used in the following evaluations are calculated by using the \gls*{eer} threshold of the entire dataset to make a binary decision if two embeddings belong to the same transmitter.

\subsubsection{Transferability Between SDRs}

In a deployed system it would be preferable to use a single common set of anchors between all receivers, since deriving new data for each receiver will be cumbersome.
In addition, it should be transferable to a new type of \gls*{sdr} device to simplify deployment or enable the system hardware to be upgraded.

\begin{figure}
    \centering
    \includegraphics[width=0.7\linewidth]{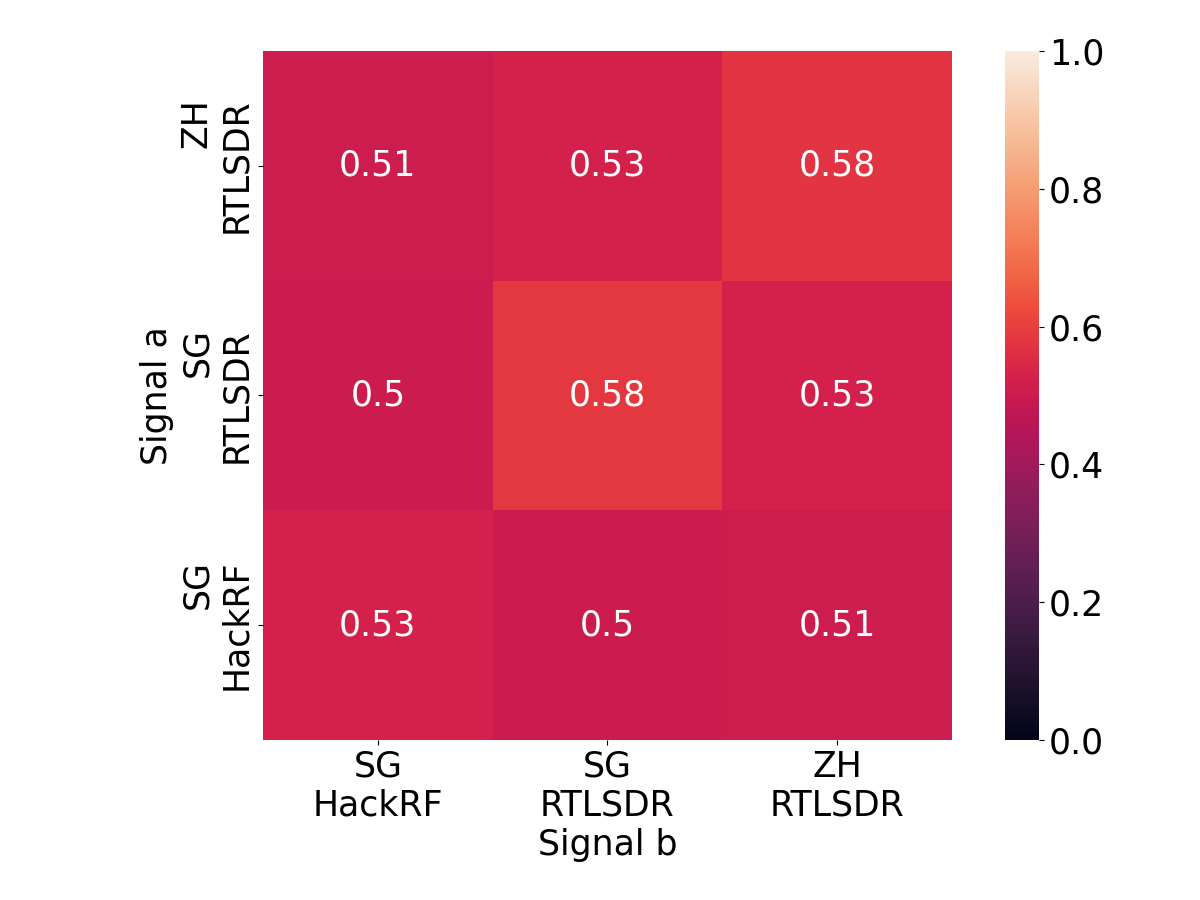}
    \caption{\gls*{auc} for different location and \gls*{sdr} combinations}
    \label{fig:accuracy-by-location-sdr}
\end{figure}

\autoref{fig:accuracy-by-location-sdr} depicts the \gls*{auc} for signals collected from different locations and \glspl*{sdr}.
It shows that while samples collected by the same \gls*{sdr} perform better, the proposed model learns to generalize beyond \gls*{sdr} boundaries.

\subsubsection{Time stability of the generated fingerprints}

\begin{table*}
\centering
\caption{Comparison of different \gls*{leo} \gls*{rff} models}
\label{tab:model-comparison}
\begin{tabular}{llll}
    \toprule
    Work                                & Target Constellation & Dataset Size                 & Best Achieved Result\\
    \midrule
    This Work                           & Orbcomm              & \datasettotalsamples{} packets & \gls*{auc} \multianchormaxauc{}, \gls*{eer} \multianchormineer{}\\
    This Work                           & Iridium              & SatIQ~\cite{WatchThisSpace} Dataset             & \gls*{auc} \iridiumauc{}, \gls*{eer} \iridiumeer{}\\
    SatIQ~\cite{WatchThisSpace}         & Iridium              & $1.7 \cdot 10^6$ ring alerts & \gls*{auc} 0.698, \gls*{eer} 0.350\\
    PastAI~\cite{PastAI}                & Iridium              & $1 \cdot 10^8$ IQ samples    & Accuracy 0.82, up to $>$0.9 (for $>$9 excluded classes) \\
    Zhu et al.~\cite{3DConvRFFIridium} & Iridium              & $1.98 \cdot 10^8$ IQ samples  & Accuracy 0.924, F1-score 0.923 \\
    \bottomrule
\end{tabular}
\end{table*}

\begin{figure}
    \centering
    \includegraphics[width=\linewidth]{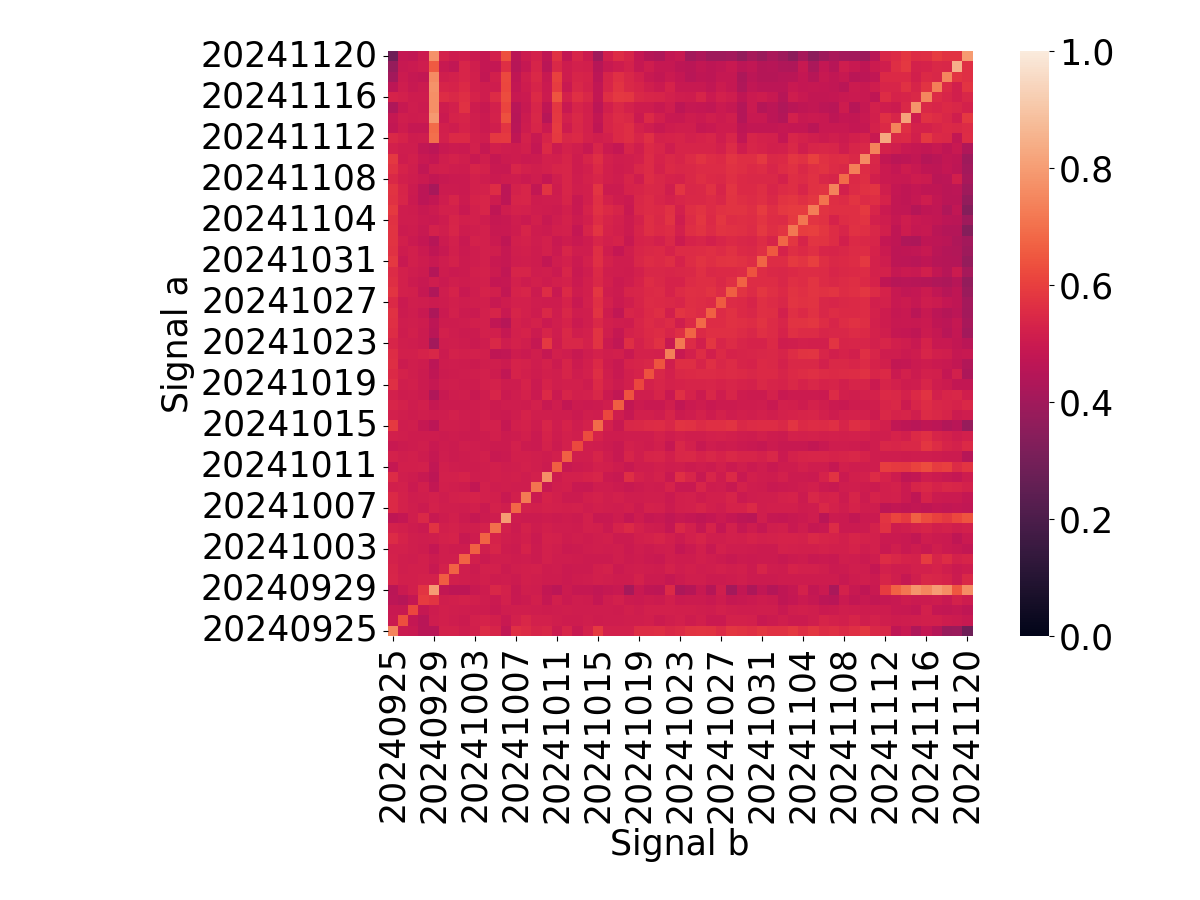}
    \caption{Impact of reception date on model performance The accuracy is calculated with the EER threshold of the model over the validation set.}
    \label{fig:accuracy-by-time}
\end{figure}
An important requirement for a deployable system is how stable the generated fingerprint is as time goes by, since a low stability requires more frequent updating of fingerprints.~\cite{cybok-plt} \autoref{fig:accuracy-by-time} shows how the reception times of the two compared signals impact the model performance. 
Over the span of roughly 2 months the performance does not show deterioration over time, except for very good performance on the same day.
The good performance of recordings on the same day is presumably that this class contains samples from the same pass that are thus likely to have experiences the same channel and atmospheric distortion and are thus more similar.

\subsubsection{Usage Time}

The fingerprinting works on a packet level.
Thus a packet consisting of 12 bytes (24 bytes for ephemeris packets~\cite{OrbcommProtocol}) has an air time of \qty{20}{ms} (\qty{40}{ms}).
The model takes an additional
\qty{30}{ms}
on a single thread of an Intel NUC with an i5-1135G7 running at \qty{2.40}{GHz} to make a prediction.

\subsection{Comparison To Other Models}

In order to test the merit of the proposed model itself, we use it to learn data published in the context of previous work on Iridium \gls*{rff}. This allows the comparison of the results achieved by the proposed model and the SatIQ model~\cite{WatchThisSpace}.

The proposed model achieves a maximal validation \gls*{auc} of \iridiumauc{} after training 100 epochs,
compared with 0.627 after training for 100 epochs as achieved by the SatIQ model. 
In order to improve the comparison, no data augmentations were used during training.

The model sizes compare as follows: SatIQ consists of around 10 M trainable parameters (\qty{38.23}{MB}) (including an autodecoder stage that is not strictly relevant for inference). This work's model contains \modelsizetrainable{} (\modelsizetrainablesize{}) trainable parameters, however, due to the slightly smaller input dimension of the Iridium data, the compared model contains only 1,079,064 (4.12 MB) trainable parameters.

On a more conceptual level, 
\autoref{tab:model-comparison} compares the results of the proposed model to previous research.

% Discussion
\section{Discussion}

% \subsection{Spoofing resistance}

One interesting point is that the performance of the model against spoofed signals is significantly better than the performance in discriminating signals from the different satellites.
We assume that this is because the task of discriminating between satellites of the same type is easier than to discriminate against a different type of transmitter (the attacker \gls*{sdr} in this case). 

%\subsection{Observations}

The model performance shows an interesting dependence on the \gls*{snr}, where the accuracy first increases but then decreases again with increasing \gls*{snr}. As mentioned in Section \ref{sec:evaluation}, the problem might be that interference caused packets to have wrongly measured \gls*{snr}. Another possibility is that due to the low occurrence of such high \gls*{snr} examples (compare \autoref{fig:dataset-stats}), the model did not learn them well.

\subsection*{Improving Model Performance}

In order to further improve the model performance, the following four approaches could be considered.

\subsubsection{Increase amount of IQ points per fingerprint}

A major difference to previous work is that only 192 \gls*{iq} samples are used to derive the fingerprint, whereas other work used 10'000~\cite{PastAI} or 11'000~\cite{WatchThisSpace}.

A comparison from the literature is a study on IEEE802.11ac \gls*{rff}~\cite{DeepLearningCNNRFFIWLAN}, which uses 128 \gls*{iq} sample points to feed a classifier network (that can distinguish 5 transmitters).
The used dataset contains 20 M samples collected from different distances. The achieved results show a good performance ($>$90\% accuracy) for close ($<$12 m) distances. The dataset stops at \qty{15}{m} distance between transmitter and receiver with an accuracy just below 75\%~\cite{DeepLearningCNNRFFIWLAN} and a decreasing trend. That study shows that fingerprinting is also possible with low-dimensional inputs but also that performance decreases with distance and, correlatedly, lower \gls*{snr}. A major difference, though, is the channel, since the cited study collected the data in an indoor environment that is more affected by multi-path effects than the typical \gls*{leo} satellite communication channel.

These comparisons indicate that a model with more input data could achieve better performance.
Studying if fingerprinting on the level of minor frames (50 packets) leads to better results would provide more insight into the impact the amount of \gls*{iq} input points have.

\subsubsection{Improve SNR}
\label{subsubsec-lowsnr}

Comparing to the literature, the captured \glspl*{snr} (compare \autoref{fig:dataset-stats}) are rather low.
An Iridium data collection campaign~\cite{PastAI} yielded
an \gls*{snr} mode of \qty{45}{dB} and 90\% of captured samples fall within \qtyrange{40}{60}{dB} \gls*{snr}.
In a simulation of \gls*{leo} satellites, the identification accuracy drastically dropped between the \qty{15}{dB} and \qty{10}{dB} simulations (from $>80\%$ down to $<40\%$ across all simulated parameters)~\cite{RFFSpikingNN}.

Another indication of \gls*{rff} performance against noise comes from a study that compared 
the performance of the SatIQ \gls*{rff} model against Jamming noise~\cite{StickyFingers}.
To compare those results the Gaussian jammer is interpreted as Gaussian channel noise instead.
At \gls*{sjr} of \qty{-0.66}{dB} (analog to \gls*{snr} of \qty{0.66}{dB}) the model on average starts to misidentify signals. These comparisons indicate that operating with higher \gls*{snr} could improve the performance.
This could be achieved by using antennae with more gain or using a preamplifier at the antenna feedpoint. 

\subsubsection{Anchor Selection}

To make a classification if a signals belongs to a satellite, currently a random anchor is used.
The performance could likely be improved by the selection of a suitable anchor.

\subsubsection{Model Training}

Training the model longer or with more data could also improve the performance of the model.
This was not attempted due to timing constraints.
More data would also enable usage of larger models before overfitting issues start manifesting.

% Future work
\section{Future work}

This work raises a number of areas for future research.

%\subsection{Further Transferability}

One point of interest would be investigating how well the model generalizes to satellites from the observed constellation that it did not encounter during training. 

Testing the model architecture on other VHF (space) communications is another near-term goal. Because the proposed model architecture does not require any particular input encoding of the \gls*{iq} samples, it could conceivably be extended to cover multiple constellations at once.

%\subsection{Vulnerability Against Adversarial Samples}

Finally, although this paper and previous works show that machine learning based \gls*{rff} systems perform well against conventional spoofing, they may potentially introduce a vulnerability against adversarial samples designed to trick the model.
Studying the vulnerability and the feasibility of injecting adversarial examples over a wireless channel may provide further insight into the security of ML-based \gls*{rff}.

\section{Conclusion}

This paper extends traditional \gls*{rff} and machine learning methods to the Orbcomm satellite constellation.
We applied new methods to deal with the constraints of the Orbcomm system and demonstrated their effectiveness.

We have also collected and provided a large dataset of Orbcomm packet captures, which will be useful for future research.
In addition, the initial assumption that training to distinguish different satellites provides good protection against spoofing attacks is validated by the much better performance of the model when used to detect spoofing compared to the baseline performance.

The final \gls*{auc} of \spoofingauc{} against packet replays over a coax connection shows the promise of the proposed model.


%\section*{Acknowledgment}




\bibliographystyle{IEEEtran}
\bibliography{IEEEabrv, ./bibliography}

\begin{thebibliography}{10}
\providecommand{\url}[1]{#1}
\csname url@samestyle\endcsname
\providecommand{\newblock}{\relax}
\providecommand{\bibinfo}[2]{#2}
\providecommand{\BIBentrySTDinterwordspacing}{\spaceskip=0pt\relax}
\providecommand{\BIBentryALTinterwordstretchfactor}{4}
\providecommand{\BIBentryALTinterwordspacing}{\spaceskip=\fontdimen2\font plus
\BIBentryALTinterwordstretchfactor\fontdimen3\font minus \fontdimen4\font\relax}
\providecommand{\BIBforeignlanguage}[2]{{%
\expandafter\ifx\csname l@#1\endcsname\relax
\typeout{** WARNING: IEEEtran.bst: No hyphenation pattern has been}%
\typeout{** loaded for the language `#1'. Using the pattern for}%
\typeout{** the default language instead.}%
\else
\language=\csname l@#1\endcsname
\fi
#2}}
\providecommand{\BIBdecl}{\relax}
\BIBdecl

\bibitem{LiveGPSSpoofingTrackerMap}
\BIBentryALTinterwordspacing
[Online; accessed 09-December-2024]. [Online]. Available: \url{https://spoofing.skai-data-services.com/}
\BIBentrySTDinterwordspacing

\bibitem{VSATInjection}
R.~Bisping, J.~Willbold, M.~Strohmeier, and V.~Lenders, ``Wireless signal injection attacks on {VSAT} satellite modems,'' in \emph{33rd USENIX Security Symposium (USENIX Security 24)}.\hskip 1em plus 0.5em minus 0.4em\relax USENIX Association, Aug. 2024.

\bibitem{WatchThisSpace}
J.~Smailes, S.~K\"{o}hler, S.~Birnbach, M.~Strohmeier, and I.~Martinovic, ``{Watch This Space}: {Securing Satellite Communication through Resilient Transmitter Fingerprinting},'' in \emph{Proceedings of the 2023 ACM SIGSAC Conference on Computer and Communications Security}, 2023.

\bibitem{3DConvRFFIridium}
S.~Zhu, Y.~Zhang, J.~Zhu, Y.~Chen, Y.~Shen, and X.~Jiang, ``3d convolution-based radio frequency fingerprinting for satellite authentication,'' in \emph{GLOBECOM 2023 - 2023 IEEE Global Communications Conference}, 12 2023, pp. 7586--7591.

\bibitem{PastAI}
G.~Oligeri, S.~Sciancalepore, S.~Raponi, and R.~D. Pietro, ``Past-ai: Physical-layer authentication of satellite transmitters via deep learning,'' \emph{IEEE Transactions on Information Forensics and Security}, vol.~18, 2023.

\bibitem{greatscottgadgetsHackRFGreat}
``{H}ack{R}{F} {O}ne - {G}reat {S}cott {G}adgets --- greatscottgadgets.com,'' \url{https://greatscottgadgets.com/hackrf/one/}, [Accessed 02-12-2024].

\bibitem{rtlsdrAboutRTLSDR}
``{A}bout {R}{T}{L}-{S}{D}{R} --- rtl-sdr.com,'' \url{https://www.rtl-sdr.com/about-rtl-sdr/}, [Accessed 02-12-2024].

\bibitem{ReviewRFF}
N.~Soltanieh, Y.~Norouzi, Y.~Yang, and N.~C. Karmakar, ``A review of radio frequency fingerprinting techniques,'' \emph{IEEE Journal of Radio Frequency Identification}, vol.~4, no.~3, pp. 222--233, 2020.

\bibitem{DeepComplexNetworksForRFF}
I.~Agadakos, N.~Agadakos, J.~Polakis, and M.~Amer, ``Chameleons' oblivion: Complex-valued deep neural networks for protocol-agnostic rf device fingerprinting,'' in \emph{2020 IEEE European Symposium on Security and Privacy (EuroS\&P)}, 09 2020, pp. 322--338.

\bibitem{WirelessDeviceIdentificationwithRadiometricSignatures}
V.~Brik, S.~Banerjee, M.~Gruteser, and S.~Oh, ``Wireless device identification with radiometric signatures,'' in \emph{Proceedings of the 14th ACM International Conference on Mobile Computing and Networking}, ser. MobiCom '08.\hskip 1em plus 0.5em minus 0.4em\relax New York, NY, USA: Association for Computing Machinery, 2008, p. 116–127.

\bibitem{DeepLearningCNNRFFIWLAN}
S.~Riyaz, K.~Sankhe, S.~Ioannidis, and K.~Chowdhury, ``Deep learning convolutional neural networks for radio identification,'' \emph{IEEE Communications Magazine}, vol.~56, no.~9, pp. 146--152, 2018.

\bibitem{MassiveExperimentalRFFStudy}
T.~Jian, B.~C. Rendon, E.~Ojuba, N.~Soltani, Z.~Wang, K.~Sankhe, A.~Gritsenko, J.~Dy, K.~Chowdhury, and S.~Ioannidis, ``Deep learning for rf fingerprinting: A massive experimental study,'' \emph{IEEE Internet of Things Magazine}, vol.~3, no.~1, pp. 50--57, 2020.

\bibitem{SatPrint}
G.~Oligeri, S.~Sciancalepore, and A.~Sadighian, ``Satprint: Satellite link fingerprinting,'' in \emph{Proceedings of the 39th ACM/SIGAPP Symposium on Applied Computing}, 04 2024, p. 177–185.

\bibitem{GPSFingerprinting}
H.~Wang, W.~Cheng, C.~Xu, M.~Zhang, and L.~Hu, ``Method for identifying pseudo gps signal based on radio frequency fingerprint,'' in \emph{2018 10th International Conference on Communications, Circuits and Systems (ICCCAS)}, 2018, pp. 354--358.

\bibitem{RobustSatAntFingerprintingRNN}
S.~Qiu, K.~Sava, and W.~Guo, ``Robust satellite antenna fingerprinting under degradation using recurrent neural network,'' \emph{Modern Physics Letters B}, vol.~36, no.~12, p. 2250043, 2022.

\bibitem{TripletLossOASIS}
G.~Chechik, V.~Sharma, U.~Shalit, and S.~Bengio, ``Large scale online learning of image similarity through ranking,'' \emph{Journal of Machine Learning Research}, vol.~11, pp. 1109--1135, 03 2010.

\bibitem{1703.07737}
A.~Hermans, L.~Beyer, and B.~Leibe, ``In defense of the triplet loss for person re-identification,'' 2017.

\bibitem{OrbcommExperience}
J.~Harms, ``The orbcomm experience,'' \url{https://connectivity.esa.int/sites/default/files/1_The_Orbcomm_Experience.pdf}, OHB Technology, Tech. Rep., 2004.

\bibitem{MultiConstellationSDRPositioning}
F.~Farhangian and R.~Landry, ``Multi-constellation software-defined receiver for doppler positioning with leo satellites,'' \emph{Sensors}, vol.~20, p. 5866, 10 2020.

\bibitem{CarrierPhaseTrackingPositioningOrbcomm}
\BIBentryALTinterwordspacing
Y.~Xie, G.~Li, H.~Qin, C.~Zhao, M.~Chen, and W.~Zhou, ``Carrier phase tracking and positioning algorithm with additional system parameters based on orbcomm signals,'' \emph{GPS Solutions}, vol.~28, no.~4, p. 184, Aug 2024. [Online]. Available: \url{https://doi.org/10.1007/s10291-024-01721-8}
\BIBentrySTDinterwordspacing

\bibitem{OppertunisticNavigationOrbcommIridium}
M.~Orabi, J.~Khalife, and Z.~Kassas, ``Opportunistic navigation with doppler measurements from iridium next and orbcomm leo satellites,'' in \emph{2021 IEEE Aerospace Conference (50100)}, 04 2021, pp. 1--9.

\bibitem{OrbcommProtocol}
\BIBentryALTinterwordspacing
M.~Kenny, ``Ever wondered what is on the orbcomm satellite downlink?'' Tech. Rep., 2002. [Online]. Available: \url{https://github.com/fbieberly/ORBCOMM-receiver/blob/master/literature/Orbcomm.pdf}
\BIBentrySTDinterwordspacing

\bibitem{RFFSpikingNN}
Q.~Jiang and J.~Sha, ``Rf fingerprinting identification based on spiking neural network for leo–mimo systems,'' \emph{IEEE Wireless Communications Letters}, vol.~12, no.~2, pp. 287--291, 2023.

\bibitem{SatelliteSpoofing}
E.~Salkield, M.~Szak\'{a}ly, J.~Smailes, S.~K\"{o}hler, S.~Birnbach, M.~Strohmeier, and I.~Martinovic, ``Satellite spoofing from a to z: On the requirements of satellite downlink overshadowing attacks,'' in \emph{Proceedings of the 16th ACM Conference on Security and Privacy in Wireless and Mobile Networks}, ser. WiSec '23.\hskip 1em plus 0.5em minus 0.4em\relax New York, NY, USA: Association for Computing Machinery, 2023, p. 341–352.

\bibitem{SatBackdooring}
T.~Zhao, N.~Wang, Y.~Wu, W.~Zhang, and X.~Wang, ``Backdoor attacks against low-earth orbit satellite fingerprinting,'' in \emph{IEEE INFOCOM 2024 - IEEE Conference on Computer Communications Workshops (INFOCOM WKSHPS)}, 2024, pp. 01--06.

\bibitem{TA-1Kreuzdipol}
\emph{TA-1 Kreuzdipol 135-152 MHz}, \url{https://www.wimo.com/media/akeneo_connector/media_files/1/8/18350_TA1_8b7b.pdf}, WiMo Antennen und Elektronik GmbH.

\bibitem{orbcomm-receiver}
F.~Bieberly, ``orbcomm-receiver,'' \url{https://github.com/fbieberly/ORBCOMM-receiver}, [Accessed 16-10-2024].

\bibitem{irfan2024reliabilityradiofrequencyfingerprintingPrePrint}
\BIBentryALTinterwordspacing
M.~Irfan, S.~Sciancalepore, and G.~Oligeri, ``On the reliability of radio frequency fingerprinting,'' 2024. [Online]. Available: \url{https://arxiv.org/abs/2408.09179}
\BIBentrySTDinterwordspacing

\bibitem{Schroff_2015}
F.~Schroff, D.~Kalenichenko, and J.~Philbin, ``Facenet: A unified embedding for face recognition and clustering,'' in \emph{2015 IEEE Conference on Computer Vision and Pattern Recognition (CVPR)}.\hskip 1em plus 0.5em minus 0.4em\relax IEEE, Jun. 2015.

\bibitem{kingma2017adammethodstochasticoptimization}
\BIBentryALTinterwordspacing
D.~P. Kingma and J.~Ba, ``Adam: A method for stochastic optimization,'' 2017. [Online]. Available: \url{https://arxiv.org/abs/1412.6980}
\BIBentrySTDinterwordspacing

\bibitem{cybok-plt}
\BIBentryALTinterwordspacing
S.~\v{C}apkun, \emph{The Cyber Security Body of Knowledge v1.0, 2019}.\hskip 1em plus 0.5em minus 0.4em\relax University of Bristol, 2019, ch. Physical Layer \& Telecommunications Security, kA Version 1.0. [Online]. Available: \url{https://www.cybok.org/}
\BIBentrySTDinterwordspacing

\bibitem{StickyFingers}
J.~Smailes, E.~Salkield, S.~Köhler, S.~Birnbach, M.~Strohmeier, and I.~Martinovic, ``Sticky fingers: Resilience of satellite fingerprinting against jamming attacks,'' in \emph{2nd Workshop on Security of Space and Satellite Systems (SpaceSec)}, Mar 2024.

\end{thebibliography}

\end{document}